\documentclass{elsart}
\usepackage{graphicx}

\begin{document}

\begin{frontmatter}

\title{Comparison of measured and simulated lateral distributions
for electrons and muons with KASCADE}

\author[fzk]{W.D. Apel},
\author[fzk]{A.F. Badea\thanksref{r1}}, 
\author[fzk]{K. Bekk},
\author[rum]{A. Bercuci},
\author[fzk,uni]{J. Bl\"umer},
\author[fzk]{H. Bozdog},
\author[rum]{I.M. Brancus}, 
\author[arm]{A. Chilingarian},
\author[fzk]{K. Daumiller},
\author[fzk]{P. Doll},
\author[fzk]{R. Engel}, 
\author[fzk]{J. Engler}, 
\author[fzk]{H.J. Gils},
\author[uni]{R. Glasstetter\thanksref{r2}}, 
\author[fzk]{A. Haungs},
\author[fzk]{D. Heck},
\author[uni]{J.R. H\"orandel}, 
\author[uni,fzk]{K.-H. Kampert\thanksref{r2}}, 
\author[fzk]{H.O. Klages},
\author[fzk]{G. Maier\thanksref{r3}},
\author[fzk]{H.J. Mathes}, 
\author[fzk]{H.J. Mayer\corauthref{corr}},
\author[fzk]{J. Milke},
\author[fzk]{M. M\"uller}, 
\author[fzk]{R. Obenland},
\author[fzk]{J. Oehlschl\"ager}, 
\author[fzk]{S. Ostapchenko\thanksref{r4}}, 
\author[rum]{M. Petcu},
\author[fzk]{H. Rebel},
\author[pol]{A. Risse},
\author[fzk]{M. Risse},
\author[uni]{M. Roth},
\author[fzk]{G. Schatz},
\author[fzk]{H. Schieler}, 
\author[fzk]{H. Ulrich},
\author[fzk]{J. van Buren},
\author[arm]{A. Vardanyan},
\author[fzk]{A. Weindl},
\author[fzk]{J. Wochele}, 
\author[pol]{J. Zabierowski}

\address[fzk]{Institut f\"ur Kernphysik, Forschungszentrum Karlsruhe,
76021~Karlsruhe, Germany}
\address[rum]{National Institute of Physics and Nuclear Engineering, 
7690~Bucharest, Romania}
\address[uni]{Institut f\"ur Experimentelle Kernphysik, Universit\"at 
Karlsruhe, 76021~Karlsruhe, Germany}
\address[arm]{Cosmic Ray Division, Yerevan Physics Institute, 
Yerevan~36, Armenia}
\address[pol]{Soltan Institute for Nuclear Studies, 90950~Lodz, 
Poland}

\corauth[corr]{corresponding author, {\it E-mail address:}
mayer@ik.fzk.de}
\thanks[r1]{on leave of absence from Nat.\ Inst.\ of Phys.\ and 
Nucl.\ Engineering, Bucharest, Romania}
\thanks[r2]{now at Fachbereich Physik, Universit\"at Wuppertal, 
42097~Wuppertal, Germany}
\thanks[r3]{present address: University of Leeds, Leeds LS2 9JT, UK}
\thanks[r4]{on leave of absence from Moscow State University, 
119899~Moscow, Russia}

\begin{abstract}
Lateral distributions for electrons and muons in extensive air 
showers measured with the array of the KASCADE experiment 
are compared to results of simulations based on the 
high-energy hadronic interaction models QGSJet and SIBYLL. 
It is shown, that the muon distributions are well described by 
both models. Deviations are found for the electromagnetic component, 
where both models predict a steeper lateral shape 
than observed in the data. 
For both models the observed lateral shapes of the electron component
indicate a transition from a light to a more heavy composition of the 
cosmic ray spectrum above the knee. 
\end{abstract}

\end{frontmatter}


\section{Introduction}

The particle lateral distribution of extensive air showers (EAS) 
is the key quantity for cosmic ray ground observations, 
from which most shower observables are derived. 
The interaction cascade, which is initiated by a high energy 
cosmic ray particle in the atmosphere, creates a multitude of 
secondary particles, which arrive nearly at the same time but 
distributed over a large area perpendicular to the direction 
of the original particle.
The disc of secondary particles may extend over several hundred
meters from the shower axis, with maximum density in the center of the 
disc, which is called the shower core. Apart from the arrival times, the 
density distribution of particles within the shower disc contain all 
informations on the primary particle, which are left after it has undergone
a millionfold multiplication process by the atmosphere. However, it is this 
multiplication process, that foremost offers the chance to observe cosmic 
rays in the ultra and very high energy region at all: 
Due to their low flux measurements at ground, carried out with 
large arrays of individual detectors which take samples of the shower 
disc at several locations, are still the only possible 
way to study these high-energy cosmic particles \cite{haungs}. \\

The lateral distributions of electrons and muons in EAS not only contains
information on the nature of the primary cosmic ray particle, which is 
related to astrophysical questions, but it also carries information relevant 
to particle physics. While the electromagnetic interactions are thought to 
be well understood, this is not true for the high energy hadronic 
interactions. The energy range and the kinematical region in which the 
first hadronic interactions of the shower development occur are far 
beyond the accessible realm of todays accelerator experiments.
Uncertainties in the description of hadronic interactions therefore imply
uncertainties in the prediction of the shape of the lateral distributions
\cite{drescher}. \\

A parameter, commonly used to describe the form of the lateral density 
distribution, is the lateral form parameter in the Nishimura-Kamata-Greisen 
(NKG)-function \cite{nishi,kamat,greis}, usually 
called \it age. \rm The name expresses the relation 
between the lateral shape of the electron distribution and the height of the 
shower maximum. Due to the statistical nature of shower development, 
the height of the shower maximum is subject to strong fluctuations.
Showers, which have started high in the atmosphere show a flat lateral 
electron distribution, as electrons in the electromagnetic cascade suffer 
more from multiple scattering processes. Such showers are called old 
and are characterised by a large value of the age parameter. 
Young showers have started deeper in the atmosphere and
had their maximum more close to observation level. This results in a 
steeper lateral electron distribution, which corresponds to a smaller 
value of the age parameter. Apart from fluctuations, the 
height of the shower maximum depends on energy and mass of the shower 
initiating primary. Therefore, the lateral shape parameter is also 
sensitive to the mass of the primary.\\

The mutual interrelation of several independent parameters on which the 
form of the lateral shape depends, makes it a delicate task to draw
unique conclusions from the results of the measurements. Moreover, the 
interpretation of the measured raw data requires a profound understandig 
on the details of the detector response functions. Sophisticated 
simulations of the whole event chain that is initiated 
by the first collision of an ultrahigh energy cosmic ray particle 
with an air nucleus in the upper atmosphere and ends with the registration
of electronical signals in the various detector components are a 
prerequisite to any reliable analysis of the lateral distributions of all 
particle components. \\

This view encourages to measure the secondary particle 
components separately, which from an experimental point 
of view requires several detector components to be
operated simultaneously. 
The detector array of the multi detector setup 
KASCADE (Karlsruhe shower core and array detector) \cite{kas} 
is designed to disentangle the electromagnetic, the hadronic, 
and the muon component of the shower disc. 
Lateral distributions of electrons, hadrons, and muons 
(for different muon threshold energies) in the primary energy range   
$5\cdot 10^{14}\mbox{eV} < E < 10^{17}\mbox{eV}$ as measured 
with KASCADE have already presented in a previous paper \cite{lat}.
In this paper, the lateral distributions of electrons and muons in 
EAS events as measured with the KASCADE array detectors will be 
compared with the predictions of detailed Monte Carlo calculations, 
which comprise the simulation of the full 
cascading process of EAS, the simulation of the array detector 
response and the final data reconstruction mechanisms. 
Whereas in \cite{lat} the parameterisations of the lateral
distributions were analysed for mean values only, here
the reconstruction is also performed on a single event basis. 
Special emphasis is given to investigations of the shape of 
the lateral distributions, the so-called 'lateral age', and its dependencies 
on primary energy and mass of the cosmic rays.
Contrary to \cite{lat}, the hadronic component, measured with the 
KASCADE central hadron calorimeter, will not be considered in the 
present analysis, as well as measurements from 
the additional KASCADE muon devices.\\

The paper is organised as follows:
After a brief description of the experimental 
setup an overview on the simulation methods is given. Then we 
shortly outline the data reconstruction scheme. A more detailed 
explanation is given on the method we use for the reconstruction 
of electron numbers and on function used to describe the measured 
and simulated lateral shapes. 
This is followed by the presentation of mean lateral 
distributions for muons and electrons, as measured with the KASCADE 
array and a comparison with the results from
the simulations. Then we show the results for the lateral shape of 
individual showers and its dependence on the shower observables 
electron and muon number. 
The simulations results, the data will be compared with, 
are mostly based on the hadronic interaction model 
QGSJet \cite{qgsjet}, but simulations with lower statistics based 
on the SIBYLL model \cite{sibyll} have also been performed. 
Therefore, results based on the SIBYLL model and the differences 
in the predictions of both models are discussed at the end of the
paper.

\section{The KASCADE experiment}

The KASCADE experiment is located at the site of the Forschungszentrum
Karlsruhe at an altitude of 110~m above sea level. A central hadron 
calorimeter is surrounded by a rectangular array of 252 scintillation detector
stations, equally spaced by 13~m and covering an area of 200x200~m$^2$.
In addition, there is a muon tracking detector with an effective area of 
128~m$^2$. The experiment measures the hadronic, muonic and electromagnetic 
components of extensive air showers in the energy range of 
$5\cdot 10^{14}~$eV up to $10^{17}~$eV of the primary particles. A detailed
description of the experiment can be found in \cite{kas}.\\

The 252 detector stations of the KASCADE detector array are organized in 16 
electronically independent clusters. Each cluster consists of 16 stations, 
except the
inner four clusters, where one station per cluster had to give way to the 
central detector. The stations of the inner four clusters contain four 
liquid scintillation detectors, each with an area of 0.8~m$^2$  
read out by one photomultiplier. The stations of the outer clusters contain 
two such detectors with 1.6~m$^2$ total area. All photomultiplier signals 
of a detector station are added, and the integrated charge of the 
signal is recorded, together with the time of the earliest detector hit
by a shower particle. These 
detectors are designed to measure arrival time and energy deposits of the 
electromagnetic component of the showers and are therefore referred to as 
$e/\gamma$-detectors here.\\

Additionally, the stations of the 12 outer clusters house 3.2~m$^2$ plastic 
scintillation detectors below a shielding of 10~cm lead and 4~cm of 
iron, which gives 0.23~GeV threshold for muons. Each detector is read out by 4 photomultiplier devices and in turn 
yields time and energy deposit information. Again, the sum of the multiplier 
signals is recorded together with the hit pattern and the time of the earliest 
detector hit. These detectors measure the muon component and are referred to as 
muon-detectors here.\\

The shower observables which are reconstructed with the KASCADE array data 
are core position, shower direction and the lateral distribution of 
electrons and muons. From this, the shower size, expressed as total number 
of electrons $N_e$ above 3~MeV and a lateral shape parameter will be derived. 
For the muon component only the total number of muons $N_{\mu}$ above 
100~MeV can be estimated. Due to the low muon densities, a reliable 
determination of the lateral shape parameter is in general not possible 
for single event analysis.

\section{Monte Carlo simulation}

A reliable interpretation of the data requires a detailed understanding of 
the physics of shower development, as well as a detailed knowledge of the 
detector response. The whole event chain, starting with 
the primary interaction in the upper atmosphere, followed by the 
cascading of the shower particles through the air up to the response of 
the detectors at KASCADE ground level has been simulated carefully.\\

The simulation of extensive air showers is performed with the program 
CORSIKA (version 6.156 and higher) \cite{corsika}. For the high energy 
hadronic interactions, the models QGSJet (version 01) \cite{qgsjet} and SIBYLL 
(version 2.1) \cite{sibyll} are used. 
Hadronic interactions with energies below 200~GeV are treated with the FLUKA 
code \cite{fluka}, and the electromagnetic component is treated with the EGS4 
package \cite{egs4}. The showers were simulated in the energy range from
$1 \cdot 10^{14}~$eV to $1 \cdot 10^{17}$~eV. With respect to computing time 
the distribution in energy was chosen to follow a power law with 
spectral index $\gamma=-2$. To represent different primary masses, the set 
contains equal numbers of showers for five different primary types, 
namely protons, helium, carbon, silicon and iron. The position of 
the shower cores were distributed randomly all over the array. 
For the shower directions an isotropic distribution was chosen.
Output of the program is a list of all particles 
reaching KASCADE ground level together with their coordinates, arrival 
time, 3-momenta and particle type.\\

These data are input to the KASCADE simulation program, which is based
on the GEANT3 package \cite{geant}. The simulation covers the whole 
experiment, with all its detectors modelled in great detail. All particles 
from a CORSIKA simulated shower are tracked through the detectors, the
surrounding air and the absorber materials. Secondaries, created 
in interactions with the detector materials are 
likewise followed. In case of the array stations, energy deposits and 
timing information are gathered during the tracking step and
converted into a photomultiplier signal and a signal time taking into
account the light collecting geometry of the detector as well as the
specific properties of the scintillation material. Output of this 
program, concerning the array part, are arrival times and multiplier 
signals for $e/\gamma$- and muon-detectors in exactly the same format, 
that is written by the real experiment after the calibration procedure, i.e.
the simulated data can be analysed with the standard KASCADE reconstruction 
software. A more detailed description of the KASCADE array simulation
can be found in \cite{MCS-Rec}.

\section{Reconstruction of particle density distributions and shower
parameters}

Detector simulations applied to CORSIKA shower events
have also been extensively used for the development and testing of the 
array data reconstruction algorithms and procedures. As the KASCADE array 
reconstruction scheme has already been described in several previous 
papers \cite{kas,lat,MCS-Rec}, only a brief overview will be given in this 
chapter. Those parts of the analysis chain, which are concerned with the 
reconstruction of the lateral electron distribution and its properties 
will be described in more detail, as they were subject to modifications 
applied for the present analysis.

\subsection{Reconstruction of the $e/\gamma$ component}

Shower direction and shower core position, as well as shower size and the
lateral form parameter (usually known as age parameter) are reconstructed 
from energy deposits and detector response times of the $e/\gamma$-detectors 
using an iterative procedure involving three steps.\\

In the first step a rough estimate of the shower direction, core position
and shower size is obtained using fast and robust algorithms,
which don't rely on any fit procedure. In the second step, the shower 
direction is determined more accurately, by evaluating the arrival times 
of the first particle in each detector. This yields an inclination resolution
better than 0.3 degrees for showers with $\lg N_e > 4.5$ \cite{kas}.
Then, corrections depending on core position and shower inclination 
are applied to the individual detector energy deposits. From this, 
particle numbers and corresponding particle densities are calculated
for each detector. A 4-parameter fit to the spatial distribution of the 
particle densities yields core position and shower size,    
and in addition, a lateral shape parameter of the charged 
particle density distribution. The core position resolution at this 
level of reconstruction is better than 0.3~m for showers with 
$\lg N_e > 4.5$, the shower size resolution at the percentage
level \cite{kas}. The e/$\gamma$-detector signals contain contributions 
also from the muon component, for which must be corrected for, which
is performed in the third step of the reconstruction scheme: As the analysis 
of the muon-detector data proceeds in parallel to the analysis of 
the e/$\gamma$-detector data, the total muon number $N_{\mu}$ is known 
at the time of step three from the step two muon analysis. 
Therefore, the expected muon density can be estimated individually for 
each e/$\gamma$-detector. The resulting signal contribution is then
accounted for in the detector probabilities for the Likelihood minimisation 
function and the combined lateral density distributions are fitted. 
Since the accuracy in core position is in general not further improved in 
this step, only shower size and lateral form parameter are varied, yielding 
the final values of the total electron number $N_e$, and the shape 
parameter of the lateral electron density distribution.

\subsubsection{ Reconstruction of particle densities }

Reconstruction of the lateral particle density distribution requires to 
interprete the measured energy deposits in terms of particle numbers.
After correcting for different track lengths in the scintillator due to
shower inclination, special attention must be paid to the 
$\gamma$-component. The electrons are accompanied by a multitude of
$\gamma$-particles, which fake additional electrons, because the photon 
efficiency of the scintillation detectors is roughly $10$\%.
The percentage of fake electrons strongly depends on core distance, because
the mean $\gamma/e$-ratio is a function of core distance. 
In addition, the detector efficiency for electrons decreases with increasing
core distance, as their kinetic energy distribution becomes more and more
soft with growing core distance. Moreover, both effects depend on shower
size and primary particle type.\\

A shower size dependent lateral energy correction function (LECF) has been 
derived using the Monte Carlo simulations described in chapter 3. 
This function gives the average expected energy deposit per 
\it shower electron \rm as a function of core distance and shower size 
and thereby accounts for the additional deposit due to accompanying photons 
and the dependence of the detector efficiency on the kinetic energy of the 
electrons. Dividing the measured energy deposit by the expected 
deposit per shower electron, an estimate on the number of electrons, 
hitting the detector is given.  

\begin{figure}[t]
\begin{minipage}[b]{.50\linewidth}
\centering\includegraphics[width=7.5cm]{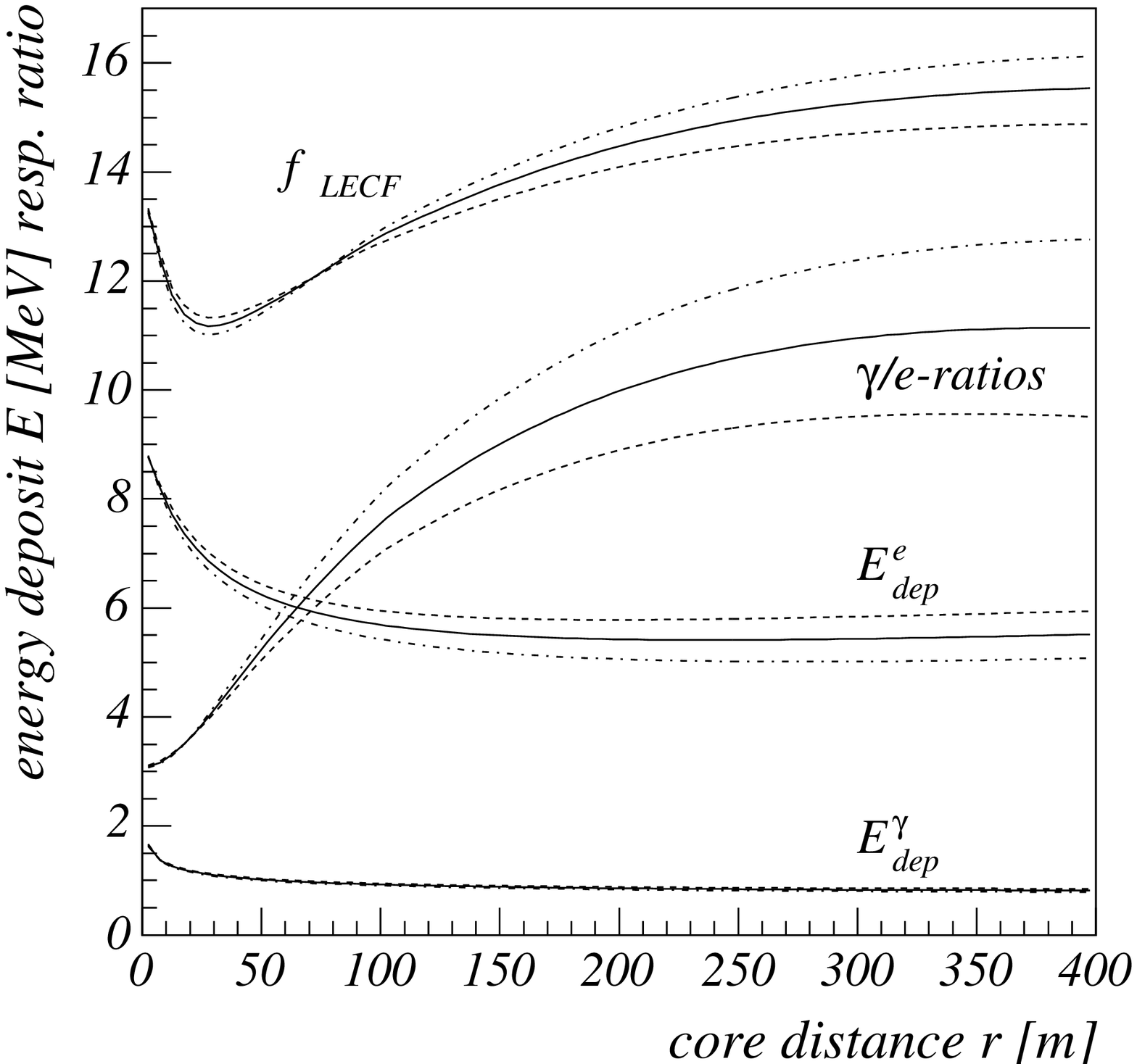}
\end{minipage}
\begin{minipage}[b]{.50\linewidth}
\centering\includegraphics[width=7.5cm]{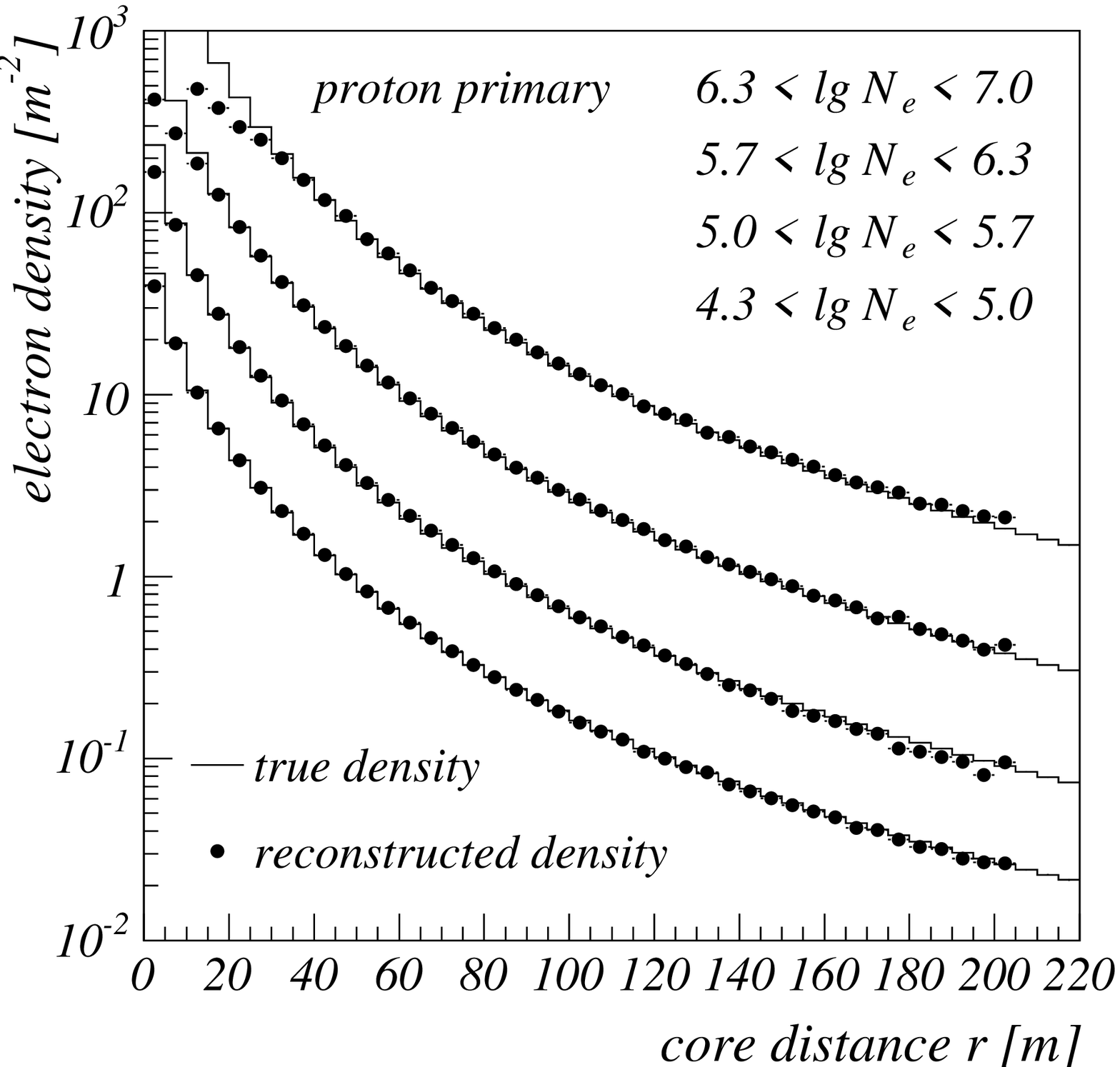}
\end{minipage}
\caption{\label{fig1}Left: Proton LECF's for $\lg N_e = 5$ (dashed lines),  
$\lg N_e = 6$ (full lines) and $\lg N_e = 7$ (dashed dotted lines). 
Also shown are corresponding $\gamma$/e-ratios 
and mean energy deposits for electrons and photons 
as parametrised and used for the LECF. Right: Mean electron
densities as reconstructed from simulated detector data for proton induced 
CORSIKA showers. The solid lines give the original CORSIKA electron 
densities. Only electrons and photons have been tracked through the 
detectors in this case.}

\end{figure}

Figure \ref{fig1} left shows
LECFs for proton induced showers for different shower sizes. The figure 
also shows the parametrisation of the mean $\gamma/e$-ratio $q_{\gamma/e}$ 
and the mean energy deposits $E_{dep}^{e}$, $E_{dep}^{\gamma}$ per electron 
and photon, respectively. The energy $E_{dep}^{tot}$ deposited in average by 
$n_e$ electrons and $n_{\gamma}$ accompanying photons in the detector is then 
given by

\begin{equation} 
E_{dep}^{tot} = n_e \cdot E_{dep}^{e} + {n}_{\gamma}\cdot E_{dep}^{\gamma},
\end{equation}

From this the functional form of the LECF ist derived according to

\begin{equation} 
f_{LECF} \equiv \frac{E_{dep}^{tot}}{{n}_e}= 
E_{dep}^{e} + \frac{n_{\gamma}}{n_{e}} \cdot E_{dep}^{\gamma}
\approx 
E_{dep}^e + q_{\gamma/e}\cdot E_{dep}^{\gamma}
\end{equation}

where the mean electron and photon energy deposits are assumed to depend 
only on the mean kinetic particle energies.\\

The sudden fall of the LECF starting at the shower center reflects the 
decrease of the mean kinetic energy of electrons and $\gamma$'s with 
increasing core distance. At a core distance of about 30~m, the 
corresponding loss in detector efficiency becomes compensated by the 
rising $\gamma$-electron ratio, which yields an increasing fraction of 
fake deposit due to $\gamma$-particles. At large core distances, 
$\gamma$'s fake nearly half one of the average energy deposit per electron.

Comparing LECFs calculated for proton and iron primaries, one finds them 
to differ by less than $1$\% within 100~m from the shower core. 
For larger distances, the differences  do not exceed $5$\%. So one is 
free to use a common LECF for the analysis of data, where the primary 
particle type is unknown. Moreover, the variation with shower size is 
less than 
100~m. There is a negligible dependence also on the inclination of the 
showers, as geometrical effects are corrected for in the reconstruction 
procedures.\\

The right part of Figure \ref{fig1} shows results of the reconstruction
when applied to simulated detector deposits, calculated from CORSIKA 
showers with the detector Monte Carlo, as described above. 
To check the reliability of the LECF, only electrons and 
$\gamma$-particles have been tracked in this case. The reconstructed electron 
distributions compare well to the distributions of the
CORSIKA electrons. Deviations from the original distribution are found 
only for large showers and small core distance. This however is due to 
saturation effects in the $e/\gamma$-detectors, which are included in the
detector simulation.

\subsubsection{ Reconstruction of shower parameters size and lateral shape}

A theoretically motivated function for the description of the lateral
electron density distribution $\rho(r)$ is given by the so-called 
Nishimura-Kamata-Greisen function (NKG) \cite{nishi,kamat,greis} 

\begin{equation} 
 \rho=N_e \cdot c(s) \cdot \left( \frac{r}{r_M}\right)^{s-2}
\left( 1+\frac{r}{r_M}\right )^{s-4.5},
\end{equation}

with the age parameter $s$, the Moliere radius $r_M$ and  the normalising
factor $c(s)$ 
\begin{equation} 
 c(s)=\frac{\Gamma(4.5-s)}{2 \pi r_M^2 \Gamma(s)\Gamma(4.5-2s)}.
\end{equation}

This function has been derived analytically for the case of purely
$e/\gamma$-induced air showers but is also used to describe the
lateral electron distribution of hadron induced showers. It is however known
\cite{fail_nkg_1,fail_nkg_2,fail_nkg_3}, that the NKG functions have
shortcomings in fitting measured EAS electron distribution, most 
obvious at large core distances. This deficit is usually addressed to the 
fact, that the NKG-function was derived for electromagnetic 
cascades, whereas hadron induced air showers are a superposition 
of a large number of independent electromagnetic showers.\\

In a typical KASCADE event, the detector distances to the core may extend 
up to 200~m. In case of large showers with sizes well above $\lg N_e = 6$,
which roughly correspond to a primary energy of 10~PeV,
the detectors close to the shower core become saturated and must be rejected
for the analysis. Thus, the lateral fit range differs for small and for 
large showers significantly in both, the upper and lower bound. The deviation
of the NKG-function from the true shape of the lateral electron distribution
therefore gives rise to systematic errors in the lateral shape parameter
(age $s$ in the NKG-formula), which depend on the lateral fit range, and 
thereby on shower size.\\

It has been pointed out \cite{lat}, that fixing the age parameter $s$ and 
instead varying the scale parameter $r_{0}$ (Moliere radius $r_M$ in 
the NKG-formula) 
considerably improves the fit behaviour of the NKG function. Unfortunately, 
this does work well only for mean lateral distributions constructed from 
a large number of showers. When fitting individual events, which suffer 
from large statistical and physical fluctuations between the detector 
stations, this method has proven to be significantly unstable. Especially 
for small showers it yields unreliable results.\\

\begin{figure}[t]
\begin{minipage}[b]{.50\linewidth}
\centering\includegraphics[width=7.5cm]{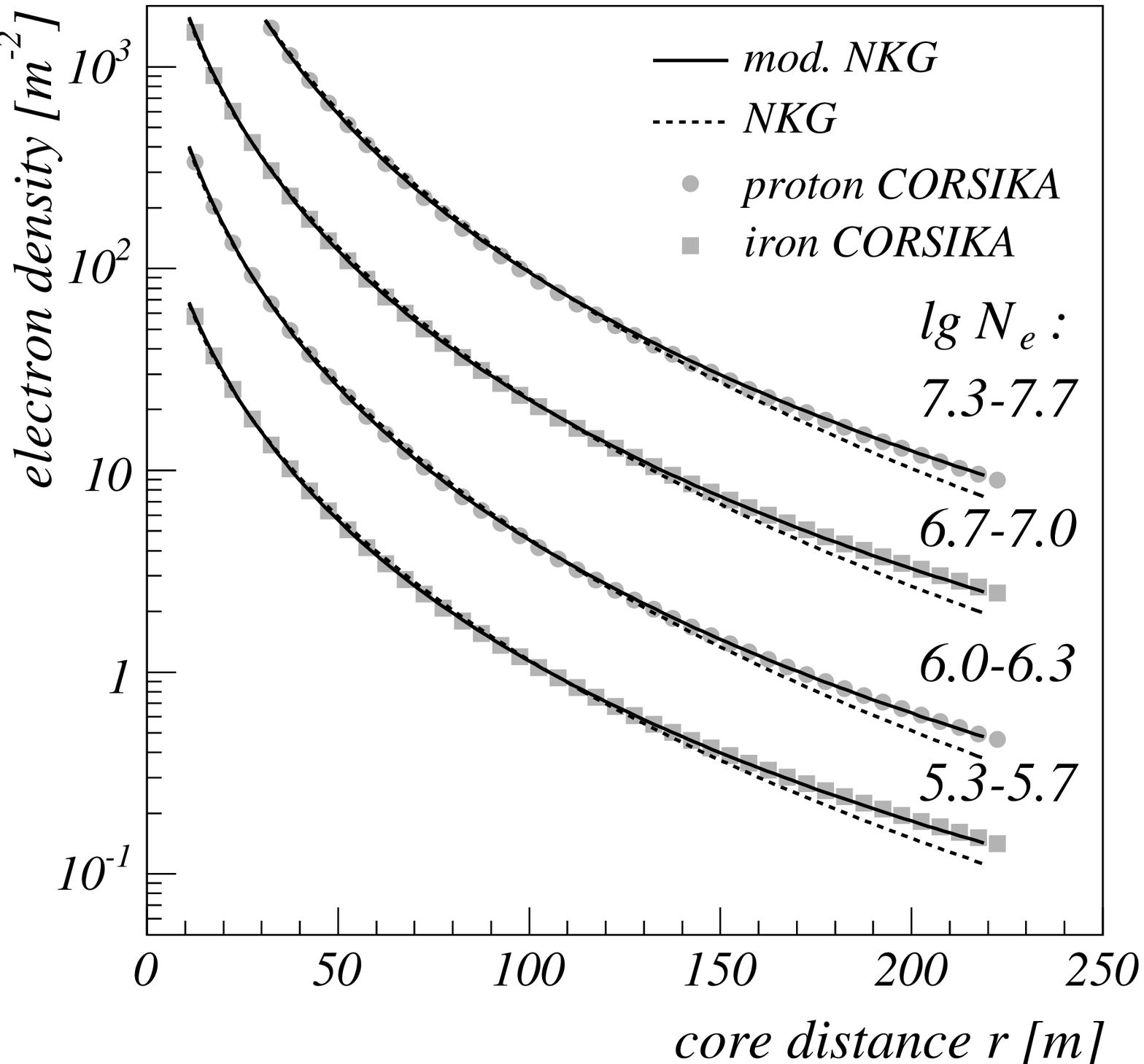}
\end{minipage}
\begin{minipage}[b]{.50\linewidth}
\centering\includegraphics[width=7.5cm]{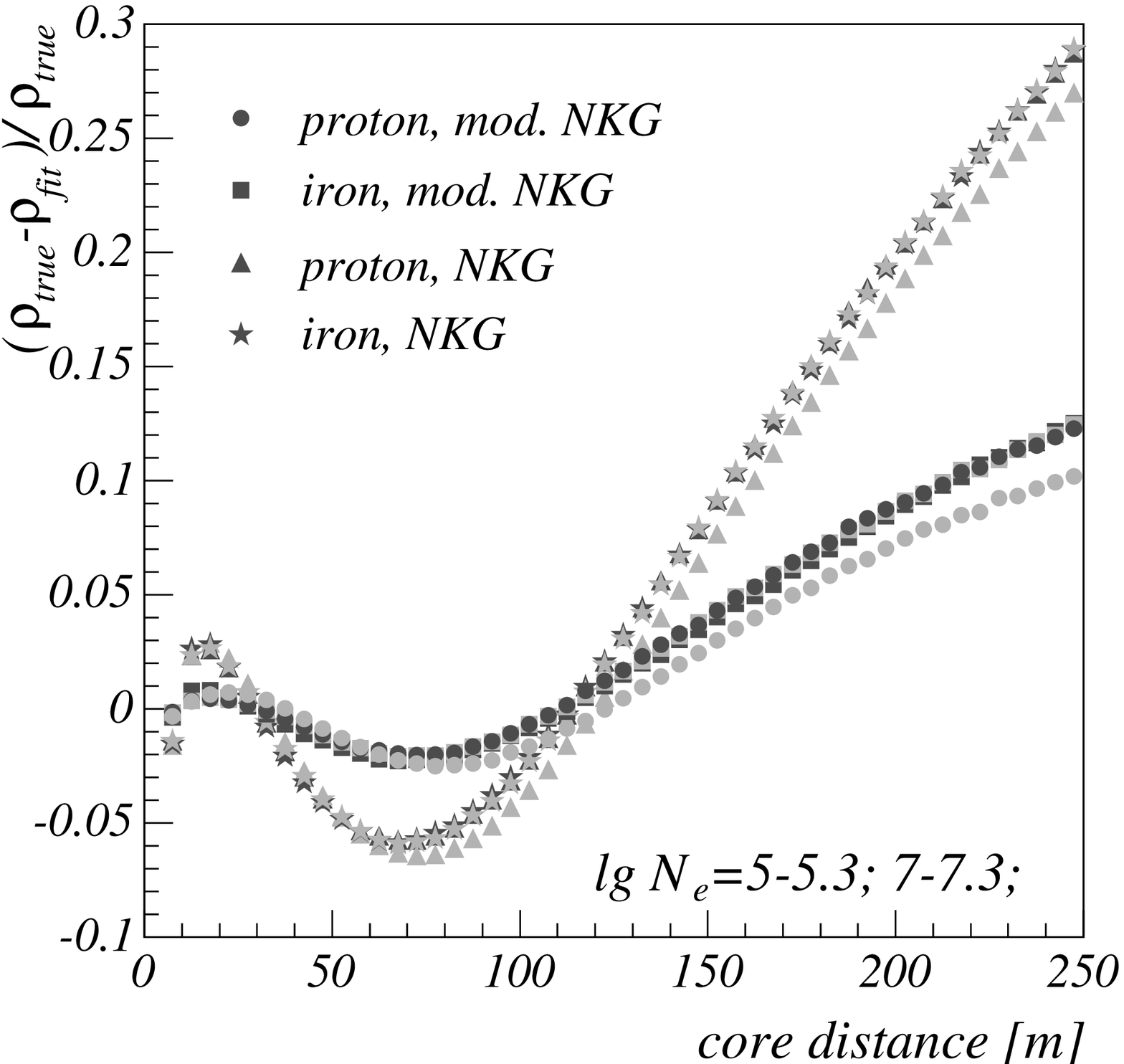}
\end{minipage}
\caption{\label{fig2}Left: Comparison of fit results for the standard NKG 
function and the modified NKG function. Shown are, as an example, fitted 
CORSIKA electron distributions (i.e. without detector simulation) for 
proton and iron showers and two different shower sizes each. 
Right: Residuals of the fitted function and the simulated CORSIKA electron
distributions for proton and iron induced showers and two different 
shower sizes (dark symbols represent the small, grey symbols the 
large $N_e$-bin). The iron symbols mostly overlap.}
\end{figure}

Other ways to cure the defects of the NKG function for describing 
electron lateral distributions of hadron induced air showers, is to
replace it by a different function (e.g. as in \cite{cap-dev}) or 
to modify its functional form by changing the values of the exponents. 
Indeed, this gives a better adaption to the shape of the lateral density 
distribution. For this, we replace equation (2) by 

\begin{equation} 
 \rho=N_e \cdot \tilde{c}(s) \cdot \left( \frac{r}{r_0}\right)^{s-\alpha}
\left( 1+\frac{r}{r_0}\right )^{s-\beta},
\end{equation}

with 
\begin{equation} 
 \tilde{c}(s)=\frac{\Gamma(\beta-s)}
 {2 \pi r_0^2 \Gamma(s-\alpha+2)\Gamma(\alpha+\beta-2s-2)}.
\end{equation}

\begin{figure}[t]
\begin{minipage}[b]{.50\linewidth}
\centering\includegraphics[width=7.5cm]{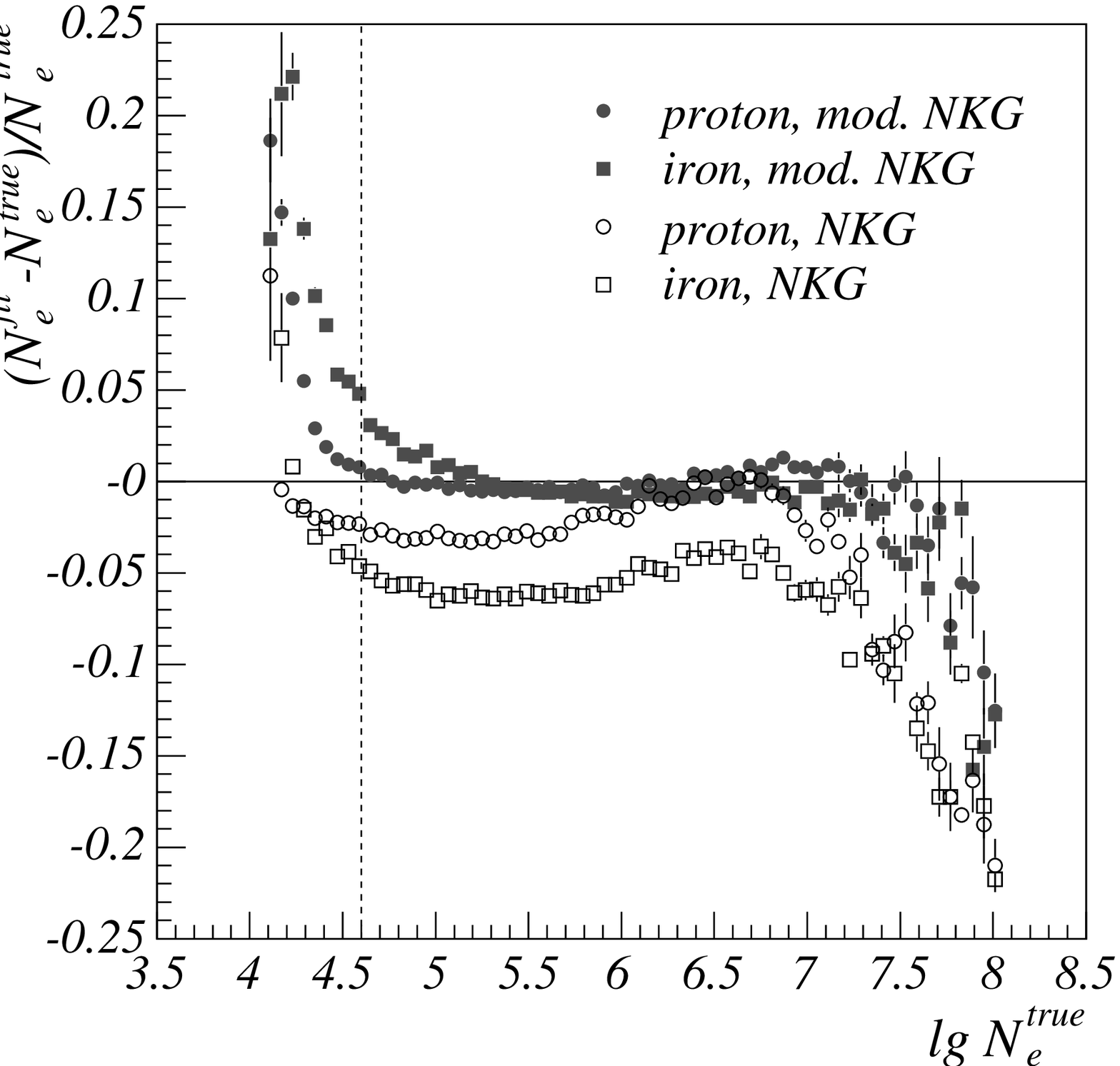}
\end{minipage}
\begin{minipage}[b]{.50\linewidth}
\centering\includegraphics[width=7.5cm]{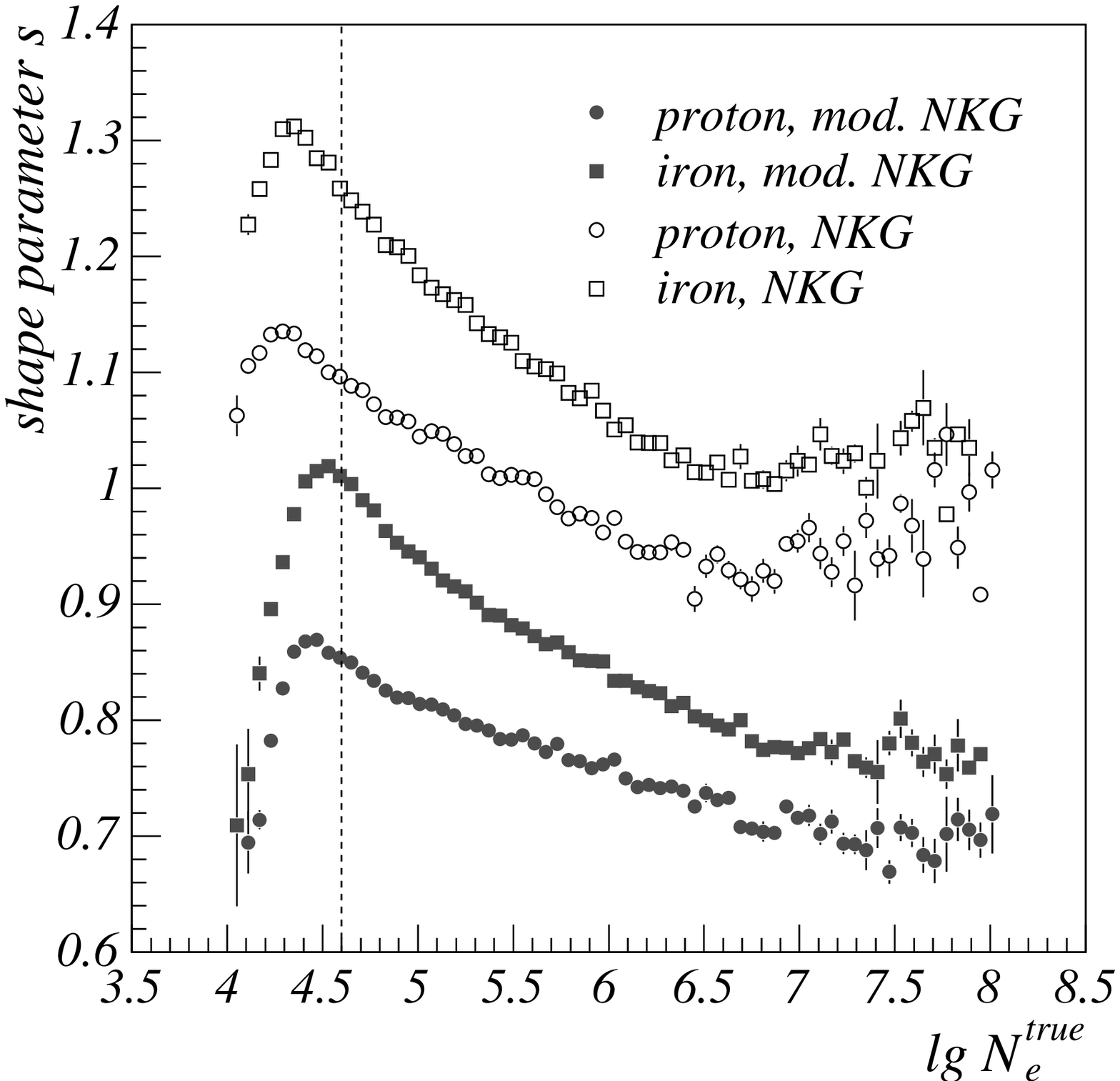}
\end{minipage}
\caption{\label{fig3}Left: Deviation of reconstructed from true shower size
for NKG and modified NKG function, when fitting individual showers. 
Right: Reconstructed shape parameter as a function of shower size using NKG and 
modified NKG function. The rise of the age parameter for showers with $\lg N_e >6.5$ 
is an artefact of the shortcomings of the NKG function in describing the lateral
distributions at large core distance. The vertical dashed lines shows the
KASCADE threshold.}
\end{figure}

Testing this function with Monte Carlo data, as optimum values for 
the exponents we have found $\alpha=1.5$ and $\beta=3.6$ , when 
$r_0=40$~m for the scale parameter is used. Optimum in this case means
an almost negligible systematic uncertainty in the reconstructed 
shower size $N_e$ over the
full KASCADE range as shown in Figure \ref{fig3}. With these values 
of $\alpha, \beta$ and $r_0$, equation (5)
limits the new parameter $s$ to the range $-0.5 < s < 1.55$. At the same 
time, of course, it looses its numerical relation to 
the longitudinal development of the electromagnetic cascade, as it is often 
mentioned for its original form. In the following, the new parameter $s$
will be called the shape or form parameter of the lateral density 
distribution.\\

As an example, the left part of Figure \ref{fig2} compares both variants
of the fit function when applied to mean lateral electron distributions
derived directly (i.e. without detector simulation) from CORSIKA simulated 
proton and iron showers for two different shower sizes. The modified 
NKG-function adapts much better to the lateral shapes, which becomes most 
obvious at large core distances. This can be seen even better 
at the right part of 
Figure \ref{fig2}, which compares relative deviations of the fit function 
from the distributions for the original NKG-function and the modified 
one. Apart from the very vicinity of the shower core, the modified 
NKG-function describes the shape of the mean lateral distributions 
with significantly smaller residuals over the whole fit range. \\

The benefits of the modified NKG-function when applied to individual showers
are shown in Figure \ref{fig3}. The left part shows results for the 
estimate of the total shower size, derived by both functions.  
With the modified function, the systematic uncertainty in shower 
size is almost zero over the range $ 5< \lg N_e < 7 $ and still below
$5$\% 
between $ 4.5 < \lg N_e < 7.5$. Even more convincing, this does
not depend on the primary particle type, contrary to the results from the
original NKG function, which fits iron induced shower profiles 
with a larger systematic uncertainty than proton induced ones.\\

The increasing systematic error towards small shower sizes below
$\lg N_e=5$ in the case of iron primaries is related to the strongly 
growing $\mu/e$-ratio. For $\lg N_e =4$, the muon density becomes
comparable to the electron density even at small core distances. This makes 
it difficult to disentangle both components in the $e/\gamma$-detector analysis
and finally leads to an overestimate in shower size. For $\lg N_e >7$ on 
the other hand, more and more detectors in the vicinity of the shower core 
get saturated. This reduces the available lateral fit range and
results in a quickly growing underestimate of the shower size. 
Both effects ultimately establish the limits in the primary energy range for
this analysis to the region between $5\cdot 10^{14}~$eV and $10^{17}~$eV.\\

The right part of Figure \ref{fig3} compares the results of the original and 
the modified NKG function for the lateral shape parameter as a function of 
the reconstructed shower size. As already mentioned, the absolute value
of the shape parameter is related to the choice of the scale radius $r_0$ 
and the values taken for the exponents $\alpha$ and $\beta$ of the modified 
NKG-function and is therefore shifted to smaller 
values. Apart from that, the results of the conventional fit function 
exhibit a rise of the shape parameter for $\lg N_e  > 6.6$ due to the 
described shortcomings of the original NKG-function and the shrinking 
of the lateral fit range with growing shower size, when detectors near to 
the shower core become saturated. This artefact is absent with the new fit
function. For shower sizes below the KASCADE threshold the age distributions
are effected by the selection procedures.

\subsection{Reconstruction of the $\mu$-component}

\begin{figure}[t]
\centering
\includegraphics[width=12cm]{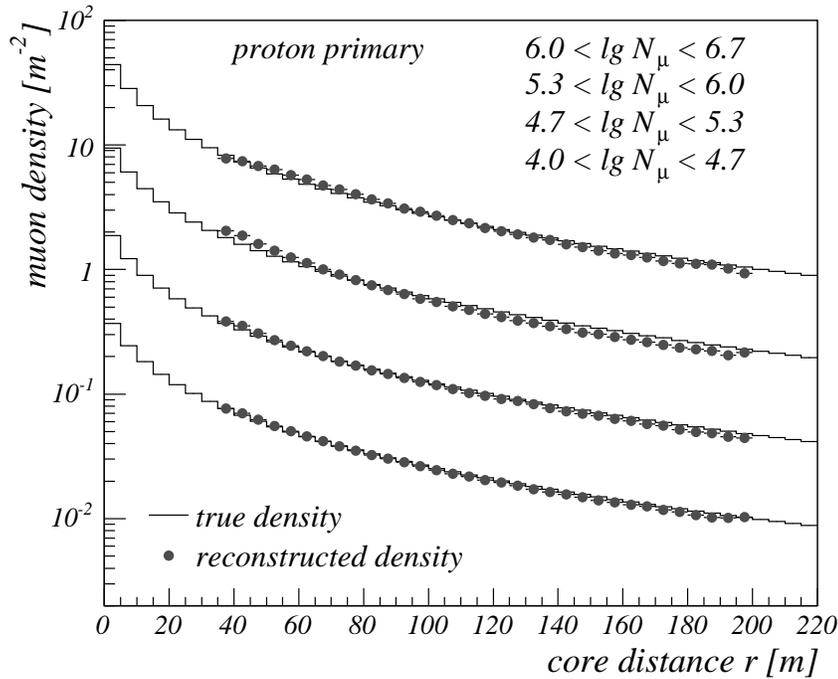}
\caption{\label{fig4} Muon lateral densities as reconstructed from
simulated muon-detector data compared to the true CORSIKA densities.}
\end{figure}

The analysis of the muon-detector signals follows the same line as
in the case of the e/$\gamma$-detectors. Again, the raw signals 
are subject to corrections due to shower inclination and core 
position using a corresponding muon LECF, which has a much simpler form than
in the e/$\gamma$-case. Additionally, corrections for 
punch through from the e/$\gamma$- and hadron component are applied 
before the muon densities for every detector are calculated. 
Detectors closer than 40~m to the shower core must be excluded from the 
analysis, because in this region punch through dominates the signal.\\

The total muon number is estimated by fitting a 
modified NKG function with exponents $\alpha=1.5$ and $\beta=3.7$. 
For the muon component this gives only a moderate improve over the original
NKG-function, which is known to fit muon lateral distributions 
already quite well, provided the scale parameter is chosen appropriately.
Here we take $r_{0\mu}=420$~m. Due to the low muon densities,
a 2-parameter fit on a single air shower basis proves unreliable. Therefore 
the total muon number
$N_{\mu}$ of the shower is estimated with a fixed lateral form parameter 
$s_{\mu}$ and a 1-parameter fit. The muon lateral shape $s_{\mu}$ is 
parametrized as a function of shower size $N_e$ from CORSIKA simulations 
and the actual value is chosen event by event during the iterative 
reconstruction. For the considered data sample $s_{\mu}$ varies between 
0.81 and 0.75, slowly decreasing with increasing shower size.\\

Reconstruction results for muon densities are illustrated in Figure \ref{fig4},
where muon distributions are compared for simulated showers with the 
corresponding true CORSIKA distributions for several ranges of the 
reconstructed total muon number $N_{\mu}$. For individual showers, 
the accuracy of this observable is moderate compared with shower size $N_e$ 
and is not better than 10 to $20$\%. This results from the poor muon 
statistics in case of small showers and punch through at large shower 
sizes. The muon size $N_{\mu}$ is input to the correction of shower size 
and shape parameter in the third step of the e/$\gamma$-detector analysis.

\section{Comparisons of KASCADE data with Monte Carlo simulations}

For the results presented here all measured showers with $\log N_e > 6$ 
have been taken into account. This sample comprises
about 170~000 events measured in a period of nearly 8 years.
Additionally, these data have been enriched with a sample of the many 
small showers, which hit the array more frequently due to the steep 
energy spectrum of cosmic rays. For this, two KASCADE runs 
with in total about 2.5 million recorded shower events were added
to the data set.\\

All showers, real or simulated, included in this analysis were subject 
to the same reconstruction procedure and to the same cuts concerning 
trigger condition, core position, inclination angle and shower shape 
parameter. Showers are restricted to
core distances less than 90~m from the array center and zenith angles less
than 30 degrees. An additional cut for showers with shape parameter value 
$s>1.4$, which is close to the upper boundary of the mathematically possible 
range, excludes showers which are frequently misreconstructed inside the 
array but actually had its core outside, or are just very small showers which
fluctuated in such a way that the reconstruction overestimated their size
by a large amount. Indeed, showers of the second kind are already very
efficiently excluded, by comparing shower sizes as reconstructed 
during step one and step two of the analysis, and cutting 
on those events, where the difference in the estimated sizes exceeds the 
expectations due to the uncertainties in both methods considerably.
Since the energy distribution of the simulated showers represents 
a spectral index of $\gamma=-2$ while the real data follow an index with 
$-2.6 < \gamma < -3.1$ fluctuations to larger values in shower size would be
less pronounced in simulations. Therefore appropriate statistical weights 
have been given to the simulated events, as will be explained below.  \\

The analysed KASCADE data set is first compared to the predictions of the 
QGSJet model, based on a sample of 1.7 Million simulated events. In addition 
a set of showers with half that statistics but based on the SIBYLL model was 
analysed (section 5.4).

\subsection{Lateral distributions of muons}

\begin{figure}[t]
\centering
\includegraphics[width=13cm]{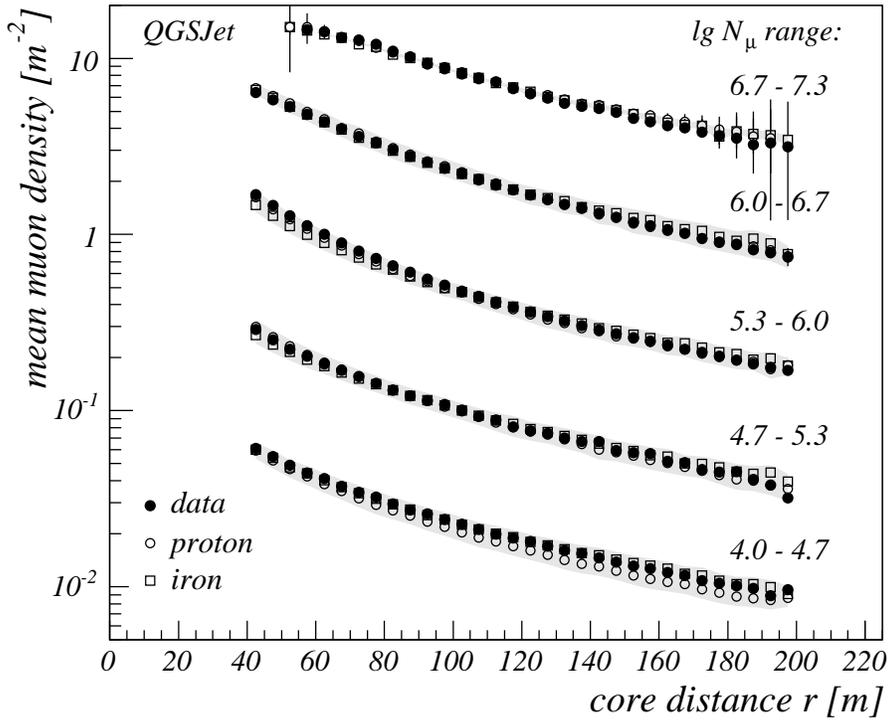}
\caption{\label{fig5} Lateral density distributions of muons as measured 
with the KASCADE array muon-detectors. The shaded bands cover the range of the 
Monte Carlo simulation results with respect to five different elemental masses
including an uncertainty in the spectral index within the range 
$-3.2 < \gamma < -2.6$. For reasons of clarity, only the results for two
elemental masses are shown additionally. For these, a spectral 
index $\gamma=-3$ is assumed.}
\end{figure}

Figure \ref{fig5} shows mean lateral distributions from the simulations 
based on the QGSJet model and compares them with the data. Showers have been 
sampled in ranges of the reconstructed total 
muon number per single event. Because of the 
steeper energy spectrum of the data, fluctuations in muon number 
would give significantly larger contributions to higher $N_\mu$-bins 
than they would for simulated events. To account for this 
effect and also to show its result on the form of the lateral
distributions, the simulated events have been analysed for a 
statistical weight representing a spectral index $\gamma=-2.6$ as well as
for weights giving an index $\gamma=-3.2$. The spectral index of the
data varies with energy, but will lie somewhere between these values.
The small shaded bands in Figure \ref{fig5} give the simulation 
results within this bounds. The lower bound of the band always corresponds to 
the larger absolute index value, i.e. to $\gamma=-3.2$. The width of the 
band for an individual primary mass is in the order of the symbol  size, only.
It is obvious, that the form of the lateral distribution 
function does only weakly depend on primary energy. Moreover,
it can be seen, that the bands of proton and iron nuclei at least 
partially overlap. The simulations therefore predict, that the 
form of the muon distribution is not sensitive to the nature 
of the primary particle. At low energies, proton induced showers 
show slightly steeper lateral shapes compared to iron 
primaries, but these differences vanish for higher energies. 
Even for small showers the differences in the density distributions  
for showers of either type do not exceed ten percent.

The lateral distributions derived from the data agree quite well with the 
Monte Carlo predictions. The simulated distributions describe the measurements
over the full KASCADE range of core distances and primary energies.
The figure thereby shows clearly, why the muon measurements at KASCADE 
are sensitive to the primary energy, but give no valuable information 
on the elemental composition of cosmic rays. This is also found by using the
SIBYLL model.

\subsection{Lateral distributions of \it charged particles \rm}

\begin{figure}[t]
\centering
\includegraphics[width=13cm]{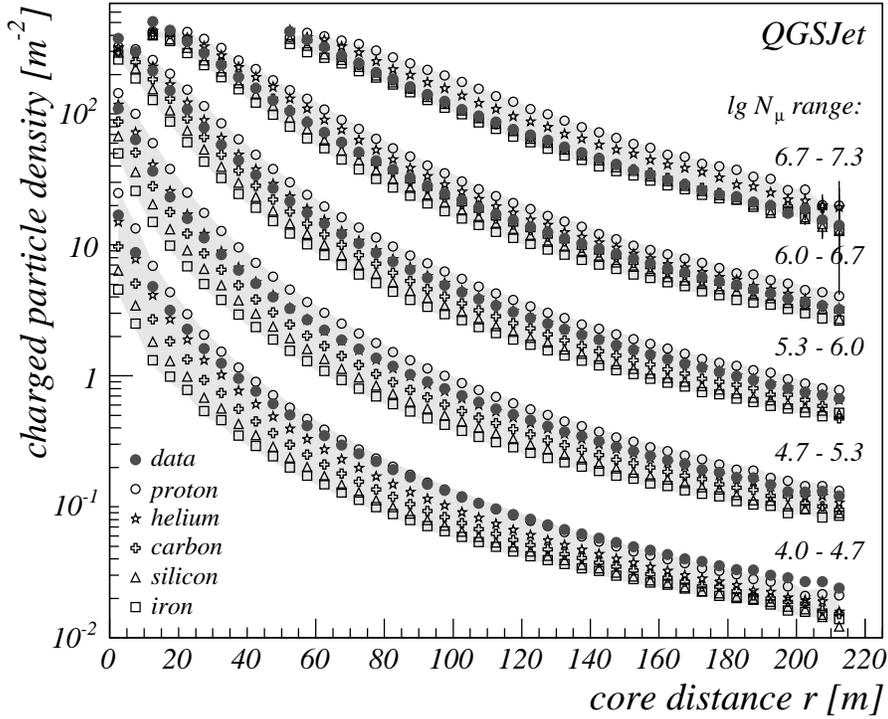}
\caption{\label{fig6} Lateral density distributions of electrons including
muons as measured with the KASCADE $e/\gamma$-detectors and by simulations. 
The shaded bands 
cover the range of the simulation results with respect to five different 
elemental masses including an uncertainty in the spectral index within 
the range $-3.2 < \gamma < -2.6$. The results for each elemental mass, 
which are also shown, assume a spectral index $\gamma=-3$. }
\end{figure}

Figure \ref{fig6} presents lateral distributions of charged particles 
as measured by the e/$\gamma$-detectors together with distributions 
derived from simulations. Though the bulk of particles are electrons, 
also muons contribute to the e/$\gamma$-detector signals, while 
contributions of photons are corrected by the LECF, and contaminations 
by hadrons are negligible. Therefore, the distributions presented here 
contain besides electrons also muons. All showers have been grouped according 
to their reconstructed total muon number.

The shaded bands of Figure \ref{fig6} again indicate the range of 
uncertainty, which 
results, when the spectral index is varied between $-3.2 < \gamma < -2.6$ 
and the primary mass from proton to iron. Again, the lower bounds
corresponds to the larger absolute value $\gamma=-3.2$, the upper
bounds to $\gamma=-2.6$. In addition, within each band the lateral 
distributions for five different primary masses are shown, assuming a spectral 
index of $\gamma=-3$ for each. The curve close to the upper bound of each band
now results from the lightest primary, the one close to the lower bound 
belongs to iron. This expresses the well known fact \cite{ap-16}, that 
showers originating from light primaries are more electron rich at sea level
than showers induced by heavy nuclei. The $e/\mu$-ratio is 
therefore a common starting point for the analysis of 
the chemical composition of cosmic rays. Furthermore, the figure 
illustrates, that showers induced by light primaries are predicted to
exhibit a slightly steeper electron distribution than showers stemming 
from heavy nuclei. \\

Comparing real data with simulations, one sees that small showers 
with $\lg N_{\mu} \approx 4$ fit simulated proton and helium 
distributions quite well, while large showers with $\lg N_{\mu} \approx 7$ 
are best described by the silicon and iron distributions, i.e. by heavier
primary particles. The figure visualizes the known \cite{kas,emu-kas} 
variation in the average $e/\mu$-ratio with shower size, 
which indicates, within the scope of the QGSJet model, a transition of the 
primary particle composition from light elements at energies below 
to heavy elements at energies above the knee. A detailed analysis of the 
electron-muon-number frequency spectrum as measured by KASCADE is subject of
a separate paper, focused on the chemical composition of cosmic rays 
\cite{holger}.\\

A closer look to Figure \ref{fig6} also reveals discrepancies between the
data and the QGSJet Monte Carlo predictions. At low energies, i.e. small muon 
numbers, the shapes of the measured distributions are slightly flatter 
than those of the simulated proton and helium distributions. For 
the smallest showers considered here, 
the measured particle densities even exceed the expected range of densities 
at core distances beyond 120~m. This kind of shortcoming can not be 
cured by any assumption on the elemental composition and will be discussed 
in more detail in the next chapter. \\

Looking at the highest energies, where data are best described by heavy 
primary showers, simulations seem to show slightly lower densities and a
flatter lateral behaviour at small core distances. In this region however,
data suffer severely from overflows, from which additional uncertainties
result.\\

It might be worth mentioning, that the shortcomings described here must 
originate completely from processes involved in the generation and development 
of the $e/\gamma$-component. The muon component contributes at low shower 
energies considerably to the lateral distributions measured with 
the $e/\gamma$-detectors. But the form of the muon lateral distribution is
well described by the Monte Carlo simulations and therefore can not be 
drawn to explain the observed deviations.  \\

\subsection{The lateral shape parameter}

The most obvious differences between the lateral shapes of the
individual elements shown in Figure \ref{fig6} are simply the 
amplitudes of the density functions, and are 
related to different $e/\mu$-ratios. A more subtle quantity, 
which in this kind of representation is difficult to compare, concerns
the functional form or the slope of the lateral distribution.\\

\begin{figure}[t]
\centering
\includegraphics[width=13cm]{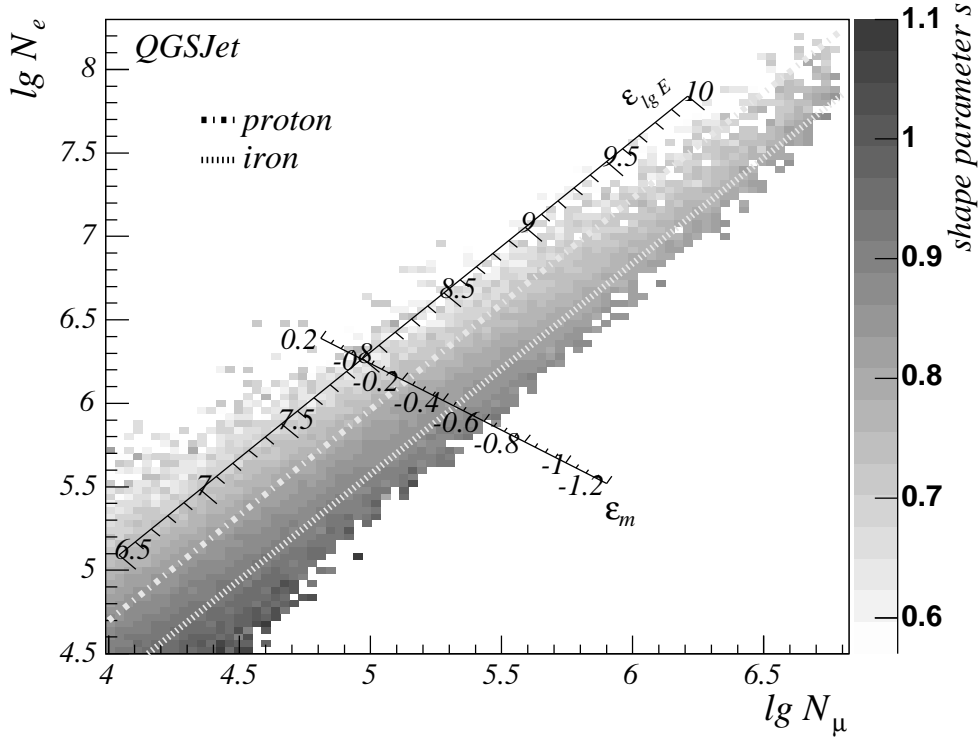}
\caption{\label{fig7} Mean shower shape parameter $s$ as a function 
of $\lg N_{\mu}$ and $\lg N_e$ for simulated showers based on the QGSJet 
model with a composition adopted from \cite{holger}. Also shown are the 
lines of maximum probability for proton and iron induced showers and the
coordinate system, used for the comparison of simulation results and data.} 
\end{figure}

The relations of shower sizes $N_e$ and $N_{\mu}$ with the shape 
parameter $s$ and with the primary mass are illustrated by Figure 
\ref{fig7}. It shows the mean reconstructed shape parameter value
as a function of the observables $\lg N_e$ and $\lg N_{\mu}$ for QGSJet 
based simulated showers. The individual events are weighted in energy 
to represent an elemental composition according to the results for the 
QGS modell as described in \cite{holger}. The lines overlaid to the distribution 
represent linear approximations to lines of maximum probability for showers 
of a single element (here shown only for proton and iron) but 
variable energy. It is obvious, that showers from light primaries 
are younger in average, i.e. have smaller shape values compared to heavy 
primaries, and showers of high energy are younger than low energy ones.\\

The lines of maximum probability in Figure \ref{fig7} offer one axis of 
a natural, rectangular chosen coordinate system to compare the data with 
the simulation results. The new coordinates are related to the 
$\lg N_{\mu} - \lg N_e - $ system by a simple rotation around the origin 
with an angle of 51,6 degrees, obtained from the simulations.
This coordinate system simply adapts the form of the event distribution in the 
$\lg N_{\mu} - \lg N_e - $ plane and will be used in the following 
to compare data and Monte Carlo results only on the basis of measured 
(or simulated) observables. While the coordinates along the lines of maximum
probability may be associated with an energy estimator $\varepsilon_{\lg E}$, 
the coordinates perpendicular to this direction measure a mass estimator 
$\varepsilon_m$. These are surely not the best possible estimators for
energy and mass but we do not want to draw quantitative conclusions from them.
Therefore the numerical values of the new coordinates are simply the values 
obtained by the rotation from the old $\lg N_{\mu} - \lg N_e - $values. \\

\begin{figure}[t]
\centering
\includegraphics[width=13cm]{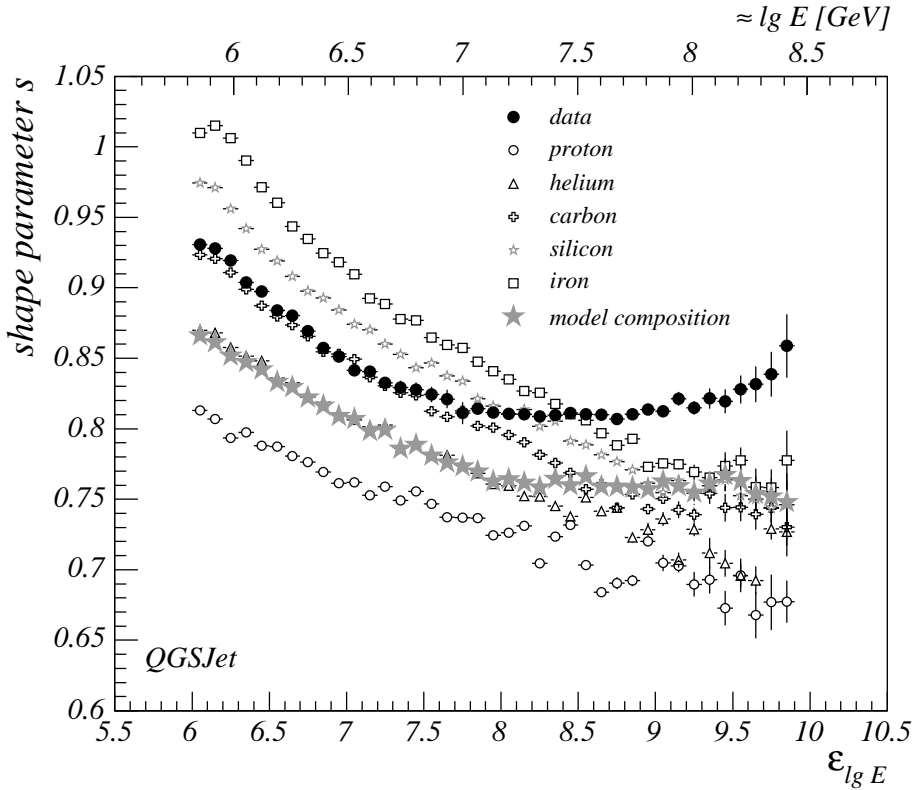}
\caption{\label{fig8} Reconstructed shape parameter 
(of the modified NKG function) 
as a function of the energy estimator $\varepsilon_{\lg E }$ for KASCADE 
data, five primary masses and a model composition as simulated using QGSJet. 
The scale on top gives a rough estimate for $\lg E$ in GeV.}
\end{figure}

The shape parameter as a function of the energy estimator 
$\varepsilon_{ \lg E}$ 
is shown in Figure \ref{fig8} for both, data and Monte Carlo simulations 
of all five elements. The Monte Carlo results confirm that showers induced by 
heavy primaries are older compared to showers of light primaries. With 
increasing energy the shape parameter value decreases for all simulated 
elements and reflects the fact, that the height of the shower maximum 
decreases with increasing energy.\\

The data fit into this picture only qualitatively. Up to an energy of about 
10~PeV, they follow the line of carbon. For low energies, this suggests 
a relatively 
heavy composition, which clearly disagrees with the predictions of figure
\ref{fig6}. For higher energies, the lateral shape parameter stays almost 
constant and crosses the line of iron at an energy of about 30~PeV. 
Beyond this crossing point, the absolute values of the measured shape 
parameter cannot be explained by any elemental composition within this Monte 
Carlo model. \\

For a more detailed investigation the data distributions are compared  
with what would be expected from the simulations, 
once a reasonable elemental composition is given. For this, the 
simulated shower events of the five elemental masses have been 
weighted with individual energy spectra, which have been reconstructed
from an analysis of the measured $N_e/N_{\mu}$-spectrum using a 
sophisticated unfolding algorithm based on the same model 
QGSJet. The resulting model composition favours light elements before
the knee and a significant contribution from heavy elements at energies
above the knee \cite{holger}. The effect on the shape parameter as a
function of the energy estimator is also shown in Figure \ref{fig8}. It 
is remarkable, that the line of the measured shape parameter values runs 
almost parallel to the line representing these adapted Monte Carlo 
predictions, but is displaced by a nearly constant amount of 
$\Delta s \sim  0.05$ over the whole energy range.\\

\begin{figure}[t]
\centering
\includegraphics[width=15cm]{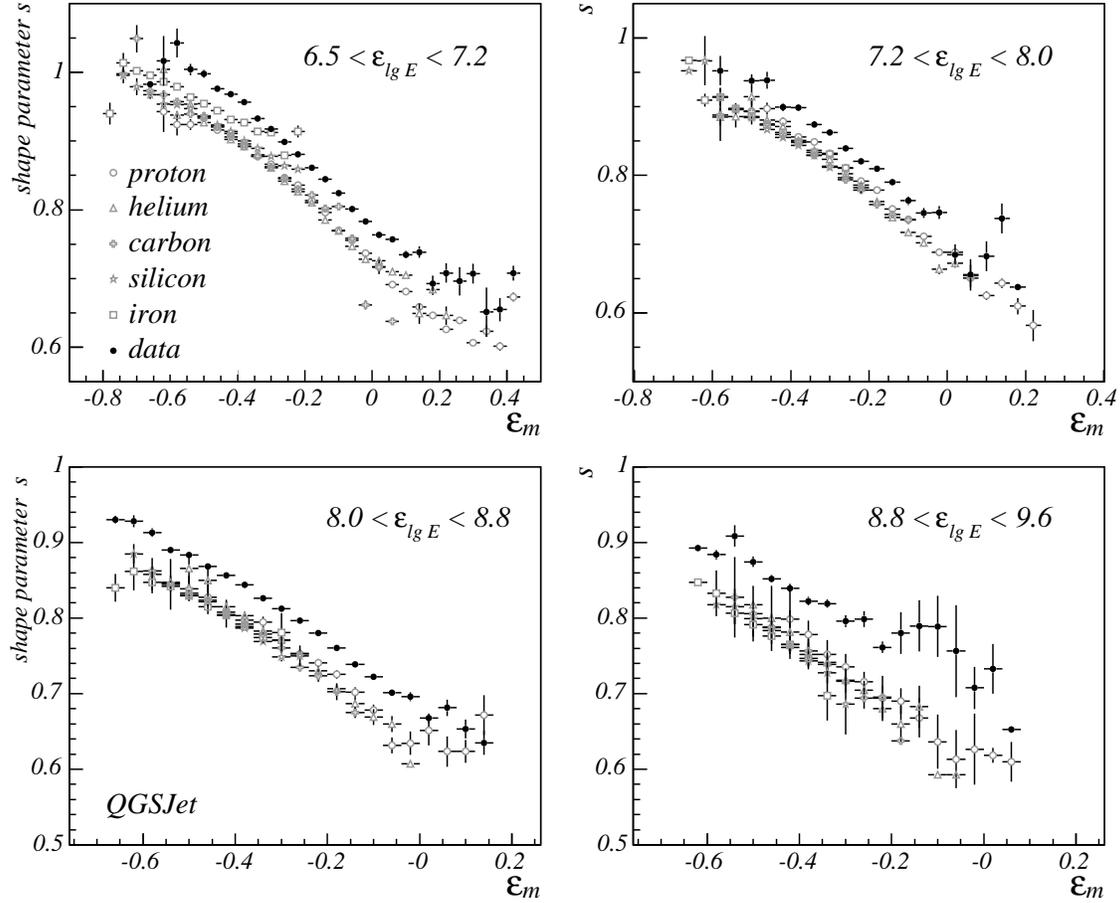}
\caption{\label{fig9} Reconstructed shape parameter (of the modified NKG 
function) as a function of
the mass estimator $\varepsilon_m$ for KASCADE data and five
primary masses as simulated using QGSJet for different ranges
of the energy estimator $\varepsilon_{\lg E }$. The line of maximum
probability for protons corresponds to a value $\varepsilon_m=-0.22$
and for iron it is $\varepsilon_m = -0.45$.}
\end{figure}

The behaviour shown in Figure \ref{fig8} may therefore be interpreted
in the same way as the form of the lateral shapes in Figure \ref{fig6}.
The almost constant value of the shape parameter for energies beyond 
10~PeV can be understood as the result of a transition from light to 
heavy nuclei in the elemental composition of cosmic rays. The offset 
between the lines of measured and simulated shape simply states, 
that the simulations in general yield slightly steeper shapes than 
observed in real showers.

This can be seen even more clearly when looking along the lines of constant
values of the mass estimator $\varepsilon_m$. This view is given in 
Figure \ref{fig9} for several ranges of the energy estimator 
$\varepsilon_{ \lg E}$, i.e. slices of Figure \ref{fig7} perpendicular 
to the lines of maximum probability. Here higher mass values correspond 
to smaller shower size and larger muon content, i.e. to showers which 
are electron-poorer.
Remarkably, all elemental masses can be seen to follow the same 
(energy dependent) functional dependence between the shape parameter 
and the mass estimator. On the other hand this may be expected, 
because showers of heavy primaries 
develop higher in the atmosphere compared to showers of light nuclei but
same energy. However, the total number of electrons present in the 
shower maximum is, for showers of the same energy, roughly independent of the 
kind of the primary nucleus. Therefore, a deeply penetrating iron 
shower may not be distinguishable in shape from a proton shower, 
which developed very early in the atmosphere. The shape of real showers 
however does not follow this functional dependence. Measured showers 
are older in average, with values deviating by an amount $\Delta s \sim 0.05$
from simulated showers, with increasing tendency at the highest energies. 
This might indicate, that real showers develop at higher altitudes 
than predicted by the simulations or/and that multiple coulomb scattering 
plays a distinct and more pronounced role in the development of the real 
electromagnetic cascade expected from simulations.

\subsection{Comparison with the SIBYLL model}

\begin{figure}[t]
\centering
\includegraphics[width=13cm]{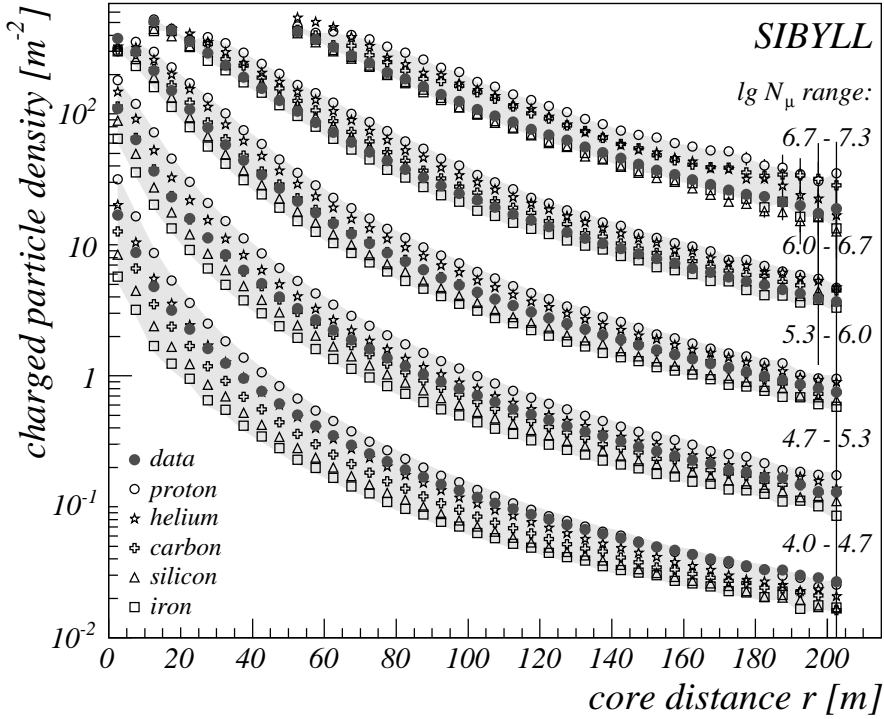}
\caption{\label{fig10} Same as Figure \ref{fig6}, but with SIBYLL 
generated showers.} 
\end{figure}

The data have also been compared with simulations based on the 
SIBYLL model. While the $e/\mu$-ratio for SIBYLL showers is larger 
in general, no notable differences to the QGSJet model were found when 
comparing the shapes of the lateral distributions of muons for equal 
total muon numbers. The lateral shapes of the electron component show 
only small differences compared to the QGSJet model, as can be seen 
from Figures \ref{fig10} and \ref{fig6}. The SIBYLL calculated shapes 
predict a more heavy composition, as a result of small differences in the 
$e/\mu$-ratios. In addition, the mean lateral electron distributions 
appear a bit younger. \\

\begin{figure}[t]
\centering
\includegraphics[width=13cm]{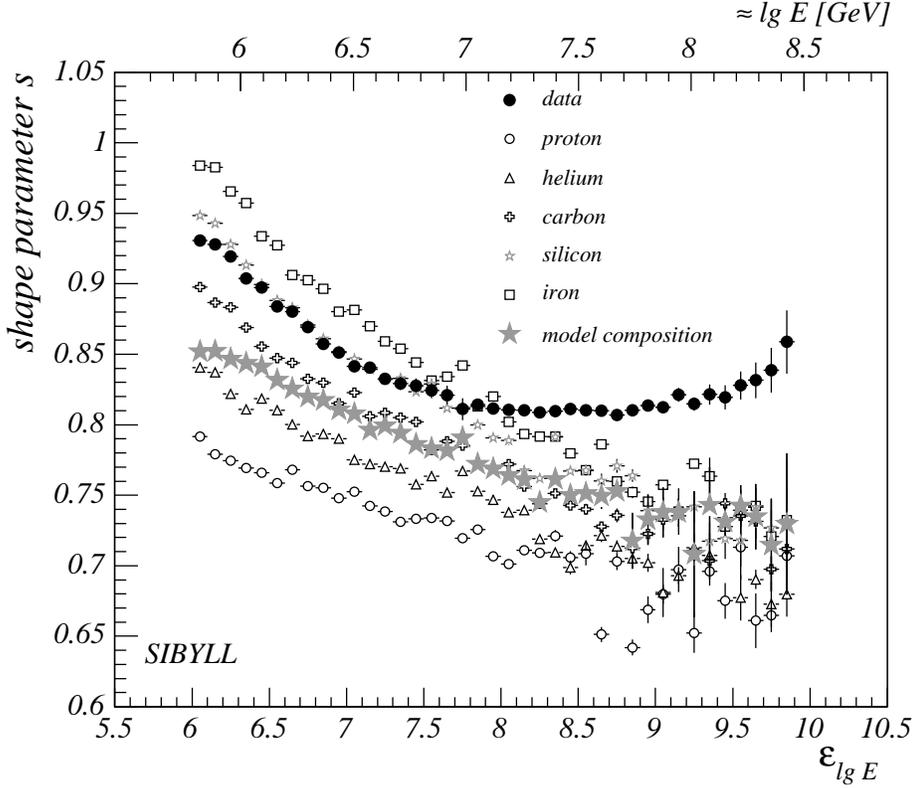}
\caption{\label{fig11} Same as Figure \ref{fig8}, but with SIBYLL generated 
showers.}
\end{figure}

Investigating the dependence of the lateral form parameter on the energy 
estimator as done in section 5.3, one can see in Figure \ref{fig11}
that SIBYLL describes
the data worse compared to QGSJet. The SIBYLL iron curve crosses the data 
already at an energy of about 10~PeV, so there is no explanation for the
measured shape values within this model for larger energies.
Comparing with Figure \ref{fig8} one finds that the mean shape of SIBYLL 
showers in general is smaller by $\Delta s \sim 0.05$ compared to QGSJet.

\begin{figure}[t]
\centering
\includegraphics[width=15cm]{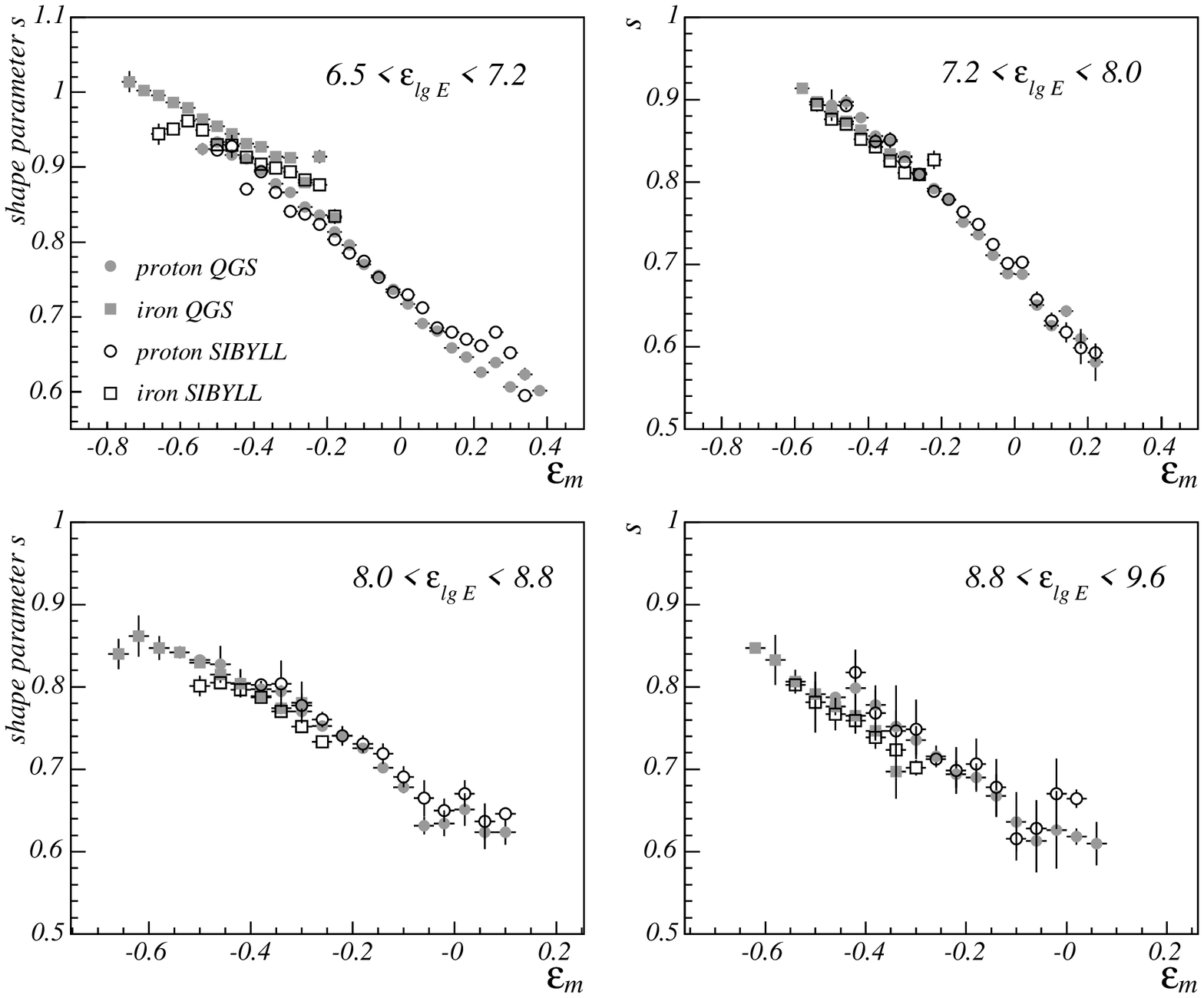}
\caption{\label{fig12} Reconstructed shape parameter 
(of the modified NKG function) as a function of 
the mass estimator $\varepsilon_m$ for QGSJet and SIBYLL based simulations
and different ranges of the energy estimator $\varepsilon_{\lg E }$. 
For reasons of clarity only two elemental masses are shown.}
\end{figure}

It may appear surprising then, that individual SIBYLL showers follow 
the same functional dependence on the mass and energy estimators 
(and therefore also on $\lg N_e $ and $\lg N_{\mu} $) than QGSJet showers 
do, as can be seen from Figure \ref{fig12}. This shows that the 
longitudinal development of the electromagnetic component must be very 
similar in both models. The difference in the mean lateral shape parameter 
is therefore simply due to a different distribution of SIBYLL events in 
the $N_e-N_{\mu}-$plane (see also \cite{holger}). SIBYLL showers are 
more electron rich and produce less muons. The abundancy maximum for a 
given primary energy is therefore shifted to lighter mass values, 
enhancing the weight of younger showers when averaging on the shape 
parameter.\\

\section{Summary and conclusions}

Lateral electron and muon density distributions of air showers as measured 
with the KASCADE array have been
compared to the results of Monte Carlo simulations based on the CORSIKA 
program using EGS4 and the two hadronic interaction models QGSJet and SIBYLL.\\

Muon lateral distributions measured with the KASCADE array muon 
detectors were found to be well described by the Monte Carlo 
simulations and  no significant differences were observed between the 
two hadronic interaction models QGSJet and SIBYLL. Moreover, muon lateral 
distributions appear very similar in shape, independent of the nature of 
the primary particle, so that details of the chemical composition can 
not show up in the comparison of data and simulations.\\

Deviations from the Monte Carlo predictions are found for the lateral 
distributions of charged particles, which were reconstructed from the 
measurements of the KASCADE e/$\gamma$-detector array.
Common to both models is that they suggest a transition from light to heavy 
nuclei in the chemical composition of cosmic rays in the energy range of 
1~PeV to  100~PeV. This is consistent with the results of an independent study
\cite{holger} based on a detailed analysis of the $\lg N_e -\lg N_{\mu}-$ 
frequency spectrum of KASCADE events.\\

Investigating in detail the shape of the lateral distributions,
the absolute values of the measured shape parameter however 
were found  to disagree
with the predictions of either of the two hadronic interaction models. 
While both models yield the same dependence of the average
shower shape on $\lg N_e $ and $\lg N_{\mu}$, the absolute values 
appear smaller than the measured shape values for the whole 
considered range in $\lg N_e$ and $\lg N_{\mu}$.\\

Looking at the mean shape parameter as a function of primary energy, 
SIBYLL yields smaller values compared to QGSJet. The reason for 
this difference between the two models 
is a kind of "lighter" distribution of events in the 
$\lg N_e -\lg N_{\mu}-$ plane in case of SIBYLL: showers of same 
primary type and energy contain on average less muons but more electrons 
and therefore make up a smaller average value for the shape parameter 
compared to QGSJet. However, also the QGSJet predictions still 
underestimate the results from measurements by an almost constant 
amount of $\Delta s \sim 0.05$.\\

Summarizing, both models are not able to describe 
the measured lateral distribution of the $e/\gamma$-component 
correctly. The details of the form of the lateral distribution depend 
on the hadronic interaction mechanism as well as on electromagnetic 
cascading processes. Thus, a variant of the QGSJet model that predicts 
a larger $e/\mu$-ratio, would give better consistency with data. 
However, the discrepancies might also be burried in the electromagnetic 
cascading algorithm EGS4 and its treatment of the multiple coulomb 
scattering process. Further improvements in the simulation models may be 
necessary, to understand and remove the discrepancies between data and 
simulations. Meanwhile, these results hopefully may help to stimulate 
this process and provide some additional hints.

\begin{ack}

The authors would like to thank the members of the engineering and technical
staff of the KASCADE collaboration, who contributed to the success of the
experiment with enthusiasm and committment. The KASCADE experiment is 
supported by the Ministry of Research of the Federal Government of Germany.
The Polish group acknowledges support by KBN research grant 1 P03B 03926 
for the years 2004-2006.

\end{ack}


\end{document}